 \documentclass[preprint, review,3p]{elsarticle}

\usepackage{amssymb}

\usepackage[utf8]{inputenc}
\usepackage{amsmath}
\usepackage{soul}
\usepackage{color}
\usepackage{graphicx}
\usepackage{array}
\biboptions{sort&compress}

\journal{Ultramicroscopy}

\begin{document}

\begin{frontmatter}

\title{A Deep Convolutional Neural Network to Analyze Position Averaged Convergent Beam Electron Diffraction Patterns}

\author[label1]{W. Xu}
\address[label1]{Department of Materials Science and Engineering, North Carolina State University, Raleigh, NC 27695, USA.}

\author[label1]{J. M. LeBeau\corref{cor1}}
\ead{jmlebeau@ncsu.edu}

\cortext[cor1]{Corresponding author}

\begin{abstract}

We establish a series of deep convolutional neural networks to automatically analyze position averaged convergent beam electron diffraction patterns. The networks first calibrate the zero-order disk size, center position, and rotation without the need for pretreating the data. With the aligned data, additional networks then measure the sample thickness and tilt. The performance of the network is explored as a function of a variety of variables including thickness, tilt, and dose.  A methodology to explore the response of the neural network to various pattern features is also presented. Processing patterns at a rate of $\sim$0.1 s/pattern, the network is shown to be orders of magnitude faster than a brute force method while maintaining accuracy. The approach is thus suitable for automatically processing big, 4D STEM data. We also discuss the generality of the method to other materials/orientations as well as a hybrid approach that combines the features of the neural network with least squares fitting for even more robust analysis.  The source code is available at https://github.com/subangstrom/DeepDiffraction.

\end{abstract}

\begin{keyword}

Machine learning \sep Convolutional neural networks \sep Position averaged convergent beam electron diffraction (PACBED) \sep Automation

\end{keyword}

\end{frontmatter}

\section{Introduction}

For a highly convergent and coherent, \aa ngstr\"om-sized electron probe, the corresponding convergent beam electron diffraction (CBED) disks strongly overlap to form a complex interference pattern. These patterns depend sensitively upon the position of the probe within the unit cell, but by averaging these patterns together, a position averaged CBED (PACBED) pattern is created \cite{LeBeau_PACBED}.  The patterns then depend strongly on sample thickness and tilt, and also reveal crystal polarity, changes in composition, octahedral distortions, and strain \cite{LeBeau_2011, Jinwoo_2012, Jinwoo_2013_PRB, Zhang_2013, Xiahan_APL, TAPLIN201669, Colin_2017}.

While PACBED patterns have been shown to be very sensitive to nanometer-level sample thickness differences and sub-milliradian tilt \cite{LeBeau_PACBED}, the patterns change with sample thickness in a non-intuitive way due to dynamical diffraction. Even so, visual inspection is often sufficient to match experimental PACBED to a library of simulated ones.  This process, however, is inherently subjective and time-consuming. To reduce human error and enhance the repeatability of the measurements, a semi-automated approach is usually employed. To this end, least square fitting (LSF) has been the primary tool \cite{Jinwoo_2012, Jinwoo_2013_PRB, Chen2015, Pollock201786, Colin_2017}. The parameter of interest, e.g.~thickness, tilt, polarity, etc., is found by searching for the best fit amongst a library of patterns. While LSF can precise and accurate \cite{Pollock201786}, it can be time-consuming to avoid local minima during the search across a broad range of parameters. Beyond processing speed, patterns alignment is an additional limiting factor. Generally, some pretreatment of the data is required by the user to locate the precise pattern center and calibrate the pixel size/scale. Specimen tilt complicates this analysis by displacing the center of intensity mass from its true position. Furthermore, alignment is further obfuscated by significant CBED disk overlap, which precludes the use of the Hough transform \cite{Klinger_2015, Pollock201786, Atherton1999795, YUEN199071}. 

Instead of brute force methods, convolutional neural networks (CNNs) have enabled breakthrough image recognition performance, even within very complex scenes \cite{Jiang2010617, CNN_bio_1, CNN_bio_2}.  For example, CNN has become the standard for applications ranging from face recognition to self-driving cars. By combining multiple, deep convolutional layers with an appropriate training set, a CNN can automatically ``learn'' high-level representations needed for robust image classification. While neural networks have shown promise for electron microscopy analysis, these powerful tools have only recently begun to be applied  \cite{KIRSCHNER200031, Pennington_2014}.   This is particularly relevant to automated PACBED analysis, as these networks have the potential to overcome many of the limitations that occur with brute force methods.  

In this work, we develop a set of deep CNNs to automatically analyze PACBED patterns, extracting pattern size, center, rotation, specimen thickness, and specimen tilt. The training and processing speeds are accelerated by the implementation of GPU calculations. Further, we show that the network architecture enables fully automatic PACBED analysis without the need for human supervision. The approach is compared to LSF using the same datasets and is found to be faster by orders of magnitude after training. Finally, we report various observations including application to 4D STEM datasets, generalizability of the trained networks to other materials, and a hybrid approach to combine neural networks with LSF for fast, robust analysis of additional parameters.


\section{Materials \& Methods}

\subsection{Experiment}

Single crystals of SrTiO$_3$, oriented along [100], and PbMg$_{1/3}$Nb$_{2/3}$O$_3$, oriented along [110], were used throughout this study. The crystals were thinned to electron transparency using wedge-polishing and low energy ion-milling using a Fischione 1050 Ar ion mill. The PMN sample was carbon coated to reduce sample charging. A probe-corrected FEI Titan G2 STEM microscope was operated at 200 kV with probe convergence semi-angle of either 13.6 mrad and 19.1 mrad.  PACBED patterns were recorded using a Gatan UltraScan 1000XP CCD camera. 

SrTiO$_3$ PACBED patterns from experiment were used to create a database for performance testing. Using the 13.6 mrad probe, a total number of 231 PACBED patterns were recorded from regions 6-120 nm thick. For the 19.1 mrad probe, a total of 156 PACBEDs were captured at thicknesses ranging from 8 to 70 nm. In both cases, the patterns exhibited random tilts up to $\sim$4 mrad. The CCD acquisition time was varied from 0.1-5 s, with a probe current of about 80 pA to incorporate varying levels of noise into the database. A 10 $\times$ 10 4D STEM dataset was collected over a 60$\times$60 nm$^2$ region of the sample with an acquisition time of 1 s/pattern. 


\subsection{Simulation}

To establish a library for neural network training, PACBED patterns were simulated using the Bloch wave method.  The Many-Beam dynamical-simulations and least-squares FITting (MBFIT) software was used for this purpose \cite{Tsuda_mbfit}. Note that the original MBFIT source code was modified to generate the PACBED output with overlapping the diffraction disks \cite{Xiahan_APL}. Patterns were calculated in 1 nm increments with thicknesses ranging from 1-120 nm at 13.6 mrad, and 1-80 nm for 19.1 mrad. At each thickness, a tilt series was also simulated with up to 4 mrad tilt along [100] and [010]. The tilts were separated by 0.25 mrad when tilt was $<$1 mrad, and 0.5 mrad otherwise. In total, 4560 and 3040 PACBED patterns were simulated for 13.6 mrad and 19.1 mrad, respectively. 

\subsection{Neural Network}

The convolutional neural networks applied here were based on the AlexNet architecture, a description of which can be found in Ref.~\cite{AlexNet}.
The network was trained via the MATLAB Neural Network Toolbox using a Titan X Pascal GPU. The network was finely tuned to measure PACBED patterns using training datasets and backpropagation through stochastic gradient descent (SGD) \cite{Bottou_2007, Zinkevich_parallel}. The learning rate in the last fully-connected layer was set to be 10 times faster than that of the other layers. To reduce overfitting and to better generalize the neural network, dropout was applied in the first two fully connected layers with a ratio of 0.5 \cite{Srivastava_2014}. Note that the original grayscale PACBED images were converted to 227$\times$227 RGB pixel images to meet AlexNet input requirements.  Although neural networks are usually considered ``black boxes'', we evaluated the regions of the PACBED pattern that created the greatest neural response using a using band-pass type mask.  The width of the annular mask was set to the size of the first AlexNet convolutional kernel size, 11$\times$11 pixels or 2.2/1.6 mrad for 13.6/19.1mrad PACBED, respectively. The pixels within the band-pass region were then set to the mean intensity of those pixels. 

\subsection{Least squares fitting}


Least square fitting was employed to benchmark the results of thickness/tilt with neural network by finding the minimum $\chi^2$ according to: 

\begin{equation}
\chi^2=\sum_{i}\sum_{j}(I_{exp,i,j}-f\cdot I_{sim,i,j})^2
\label{eq:LSF}
\end{equation}

\noindent where I$_{exp}$ and I$_{sim}$ are the experiment and simulation intensities, respectively.  The factor $f$ is included to match the intensity scale difference between the experiment and simulation. During the search of the global minimum of $\chi^2$, the simulated PACBED patterns were automatically scaled, rotated and shifted. The precision of LSF was estimated using the method suggested in Ref.~\cite{Pollock201786}. Further, when the actual specimen thickness was above $\sim$70 nm, LSF tended to converge to a thinner value. This was found to be primarily due to excess background due to inelastic scattering, which was empirically overcome by subtracting a uniform background intensity of 0.4 scaled intensity/nm from experiment.


\section{Algorithm Description}

\begin{figure*}[h!]
\centering
\includegraphics[width=6.0in]{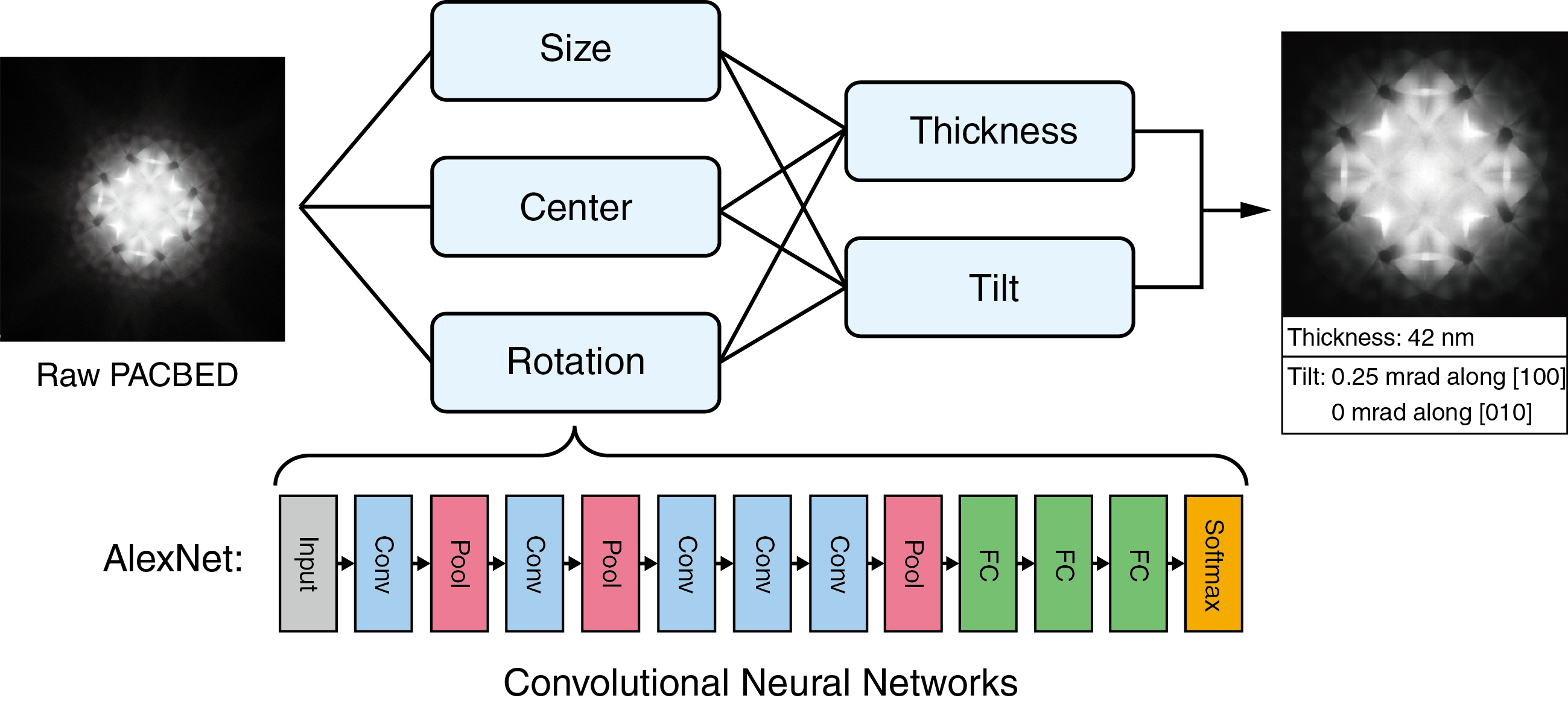}
\caption{The configuration of the convolutional neural networks used for automatic PACBED pattern measurements.}
\label{fig:PACBED_CNN}
\end{figure*}

\begin{figure}[h!]
\centering
\includegraphics[width=3.54in]{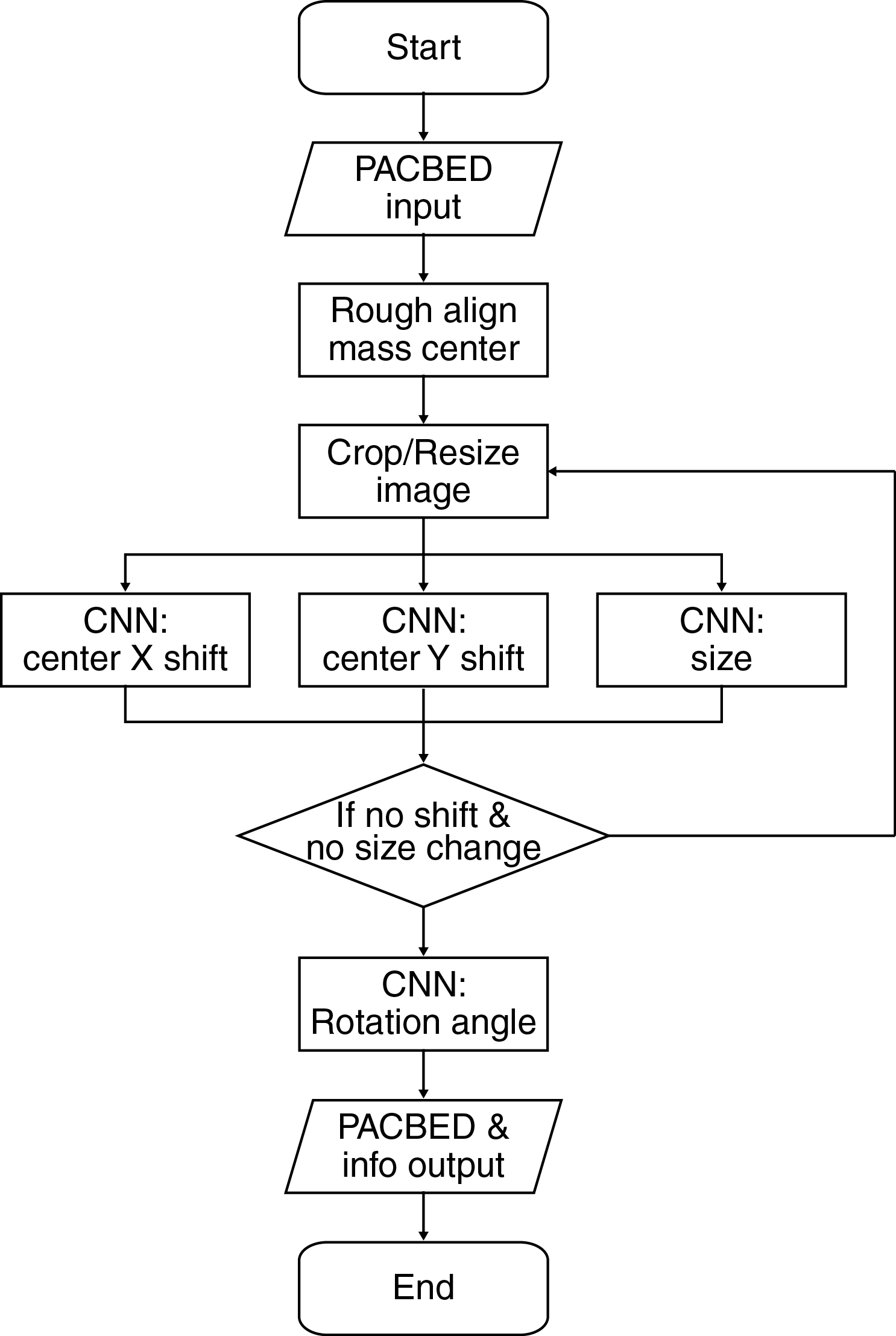}
\caption{Flow chart of the convolutional neural networks implemented during the automated alignment procedure.}
\label{fig:flow_chart}
\end{figure}

The overall neural network configuration can be seen in Figure \ref{fig:PACBED_CNN}, which contains a total number of five CNNs combined for the tasks of PACBED thickness and tilt measurement. Specifications of the implemented CNN networks are listed in Table \ref{table:training}. In the first stage, the zero order disk size, disk center, and pattern rotation angle are measured. The flow of the procedure is illustrated in Figure \ref{fig:flow_chart}. There are a number of automated procedures that are applied before passing to the CNNs. Rough estimates of the of the pattern center and size are provided by fitting the integrated intensity of the PACBED pattern along both horizontal and vertical direction to Gaussian functions.  Precision here is not essential as these variables are iterated.

The roughly centered and cropped patterns are then passed to subsequent CNNs to refine the center and shift measurements. To measure shift along both horizontal and vertical directions, the same CNN is used, but with the pattern rotated 90$^\circ$. Updated center and size variables are used to realign the original PACBED dataset until convergence. The rotation angle is then determined via another trained CNN, but without the need for iteration. Prior to thickness and tilt measurements, uniform background is subtracted from the pattern to account for the contribution from inelastic scattering. This improves the network performance, particularly when determining sample thicknesses above $\sim$70 nm. For more details, see Section \ref{sec:thicknessDependence}. As part of the process, this background subtraction value is converged while determining determining thickness and tilt. It is also important to note that only positive tilt values from 0-4 mrad along $\left[100\right]$ and $\left[010\right]$ direction are measured using the trained CNN, which is justified by the four-fold symmetry for the zone considered. The sign of tilt  along $\left[100\right]$ and $\left[010\right]$ can then be identified according to the relative image intensity in four pattern quadrants. This treatment reduces the complexity of using a CNN to measure tilt. 



\subsection{Neural network training}


AlexNet is configured with well-defined, initial convolutional kernels to extract local features for image classification. Utilizing these initial kernels not only speeds up the training process, but also improves the network performance for PACBED pattern identification. To prepare the neural network for the training, the convolution filter bias/weights and first two fully connected layers are taken directly from AlexNet. In addition, we replace the classification layers (fully-connected and softmax) with image classification categories specific to PACBED.  

The training datasets are prepared using the simulated PACBED patterns. Two key factors must be considered for successful CNN training. First, both thickness and tilt variation are included in all CNN training sets. While ideally every PACBED pattern from experiment would be captured perfectly on-zone, patterns with sub-mrad tilt are difficult to avoid in practice. The inclusion of tilt generalizes the CNN more effectively and improves measurement accuracy.  Second, and to further generalize the network, a wide of factors impacting experiment are also included in the simulated training sets through data augmentation.  These include using random affine transformations to approximate geometric distortion introduced by the projection lens aberrations, Gaussian blurring to account for inelastic scattering and the CCD point spread function, and shot noise. Random flips of the training dataset are also included.  A summary of network training for each CNN are included in Table \ref{table:training}.



\begin{table}[h!]
	\caption{Convolutional neural network training and data augmentation}
	\centering
  \centering
  \begin{tabular}{ m{3.5cm} m{3.8cm} m{4.0cm} m{1.5cm} m{1.6cm} }
    \hline
    \textbf{Neural Network} & \textbf{Target Precision \ and Application Range} & \textbf{Augmentation} & \textbf{Training \ Points} & \textbf{Learning rate} \\
    \hline
    \begin{minipage}[t]{2.5cm}
    Size \\
    (zero order disk diameter) \\
    \end{minipage}
    &
    \begin{minipage}[t]{3.8cm}
    $\pm$0.5-0.8\% size change \\
    (50-82\% image width) \\
    \end{minipage}
    & 
    \begin{minipage}[t]{4.0cm}
    distortion, rotation, shift, blur, intensity, random flip \\
    \end{minipage}
    &
    \begin{minipage}[t]{3cm}
    5.2M \\
    \end{minipage}
    &
    \begin{minipage}[t]{1.6cm}
    Start 1e-3 \\
    Final 2.5e-6 \\
    \end{minipage}
    \\ 
    \begin{minipage}[t]{3.5cm}
    Center \\
    (shift along image horizontal direction) \\
    \end{minipage}
    &
    \begin{minipage}[t]{3.8cm}
    $\pm$2 pixels \\
    (-30-30 pixels) \\
    \end{minipage}
    & 
    \begin{minipage}[t]{4.0cm}
    distortion, blur, size, intensity \\
    \end{minipage}
    &
    \begin{minipage}[t]{3cm}
    5.9M \\
    \end{minipage}
    &
    \begin{minipage}[t]{1.6cm}
    Start 2e-4 \\
    Final 2.5e-6 \\
    \end{minipage}
    \\ 
    \begin{minipage}[t]{3.5cm}
    Rotation \\
    \end{minipage}
    &
    \begin{minipage}[t]{3.8cm}
    $\pm$1$^\circ$ \\
    (-44-45$^\circ$) \\
    \end{minipage}
    & 
    \begin{minipage}[t]{4.0cm}
    distortion, shift, random crop, blur, intensity \\
    \end{minipage}
    &
    \begin{minipage}[t]{3cm}
    4.3M \\
    \end{minipage}
    &
    \begin{minipage}[t]{1.6cm}
    Start 2e-4 \\
    Final 1e-4 \\
    \end{minipage}
    \\ 
    \begin{minipage}[t]{3.5cm}
    Thickness (13.6 mrad) \\
    \end{minipage}
    &
    \begin{minipage}[t]{3.8cm}
    $\pm$1 nm \\
    (1-120 nm in 0-4.2 mrad) \\
    \end{minipage}
    & 
    \begin{minipage}[t]{4.0cm}
    distortion, rotation, shift, blur, noise, random crop, intensity, random flip, tilt variation \\
    \end{minipage}
    &
    \begin{minipage}[t]{3cm}
    37.2M \\
    \end{minipage}
    &
    \begin{minipage}[t]{1.6cm}
    5e-6 \\
    \end{minipage}
    \\ 
    \begin{minipage}[t]{3.5cm}
    Tilt (13.6 mrad) \\
    \end{minipage}
    &
    \begin{minipage}[t]{3.8cm}
    $\pm$0.25 mrad (0-1 mrad) \\ 
    $\pm$0.50 mrad (1-4.2 mrad) \\
    \end{minipage}
    & 
    \begin{minipage}[t]{4.0cm}
    distortion, rotation, shift, blur, noise, random crop, intensity, random flip, thickness variation \\
    \end{minipage}
    &
    \begin{minipage}[t]{3cm}
    35.7M \\
    \end{minipage}
    &
    \begin{minipage}[t]{1.6cm}
    5e-5 \\
    \end{minipage}
    \\ 
    \begin{minipage}[t]{3.5cm}
    Thickness (19.1 mrad) \\
    \end{minipage}
    &
    \begin{minipage}[t]{3.8cm}
    $\pm$1 nm \\
    (1-80 nm) \\
    \end{minipage}
    & 
    \begin{minipage}[t]{4.0cm}
    distortion, rotation, shift, blur, noise, random crop, intensity, random flip, tilt variation \\
    \end{minipage}
    &
    \begin{minipage}[t]{3cm}
    31.2M \\
    \end{minipage}
    &
    \begin{minipage}[t]{1.6cm}
    5e-6 \\
    \end{minipage}
    \\ 
    \begin{minipage}[t]{3.5cm}
    Tilt (19.1 mrad) \\
    \end{minipage}
    &
    \begin{minipage}[t]{3.8cm}
    $\pm$0.25 mrad (0-1 mrad) \\ 
    $\pm$0.50 mrad (1-4.2 mrad) \\
    \end{minipage}
    & 
    \begin{minipage}[t]{4.0cm}
    distortion, rotation, shift, blur, noise, random crop, intensity, random flip, thickness variation \\
    \end{minipage}
    &
    \begin{minipage}[t]{3cm}
    24.6M \\
    \end{minipage}
    &
    \begin{minipage}[t]{1.6cm}
    5e-5 \\
    \end{minipage}
    \\
    \hline
  \end{tabular}
  \label{table:training}
\end{table}



\section{Neural network performance}

The trained CNNs are tested with a series of SrTiO$_3$ PACBED patterns from experiment at different thicknesses and tilts. To evaluate the network performance, the results from the CNN measurement are  compared with the results from LSF and visual inspection. The overall performance is listed in Table \ref{table:testing}, while typical thickness and tilt classification examples are presented in Table \ref{table:exp_compare}.

\begin{table}[h!]
	\caption{Validation results comparing the convolutional neural networks to least squares fitting.}
	\centering
  \centering
  \begin{tabular}{ m{3.5cm} m{3cm} m{5cm} }
    \hline
    \textbf{Neural Network} & \textbf{Simulation} & \textbf{Experiment}\\
    \hline
    \begin{minipage}[t]{2.5cm}
    Size \\
    \end{minipage}
    &
    \begin{minipage}[t]{3.8cm}
    96.5\% \\
    \end{minipage}
    & 
    \begin{minipage}[t]{4.0cm}
    96.0\% ($\pm$1\% size change)\\
    99.4\% ($\pm$2\% size change)\\
    \end{minipage}
    \\ 
    \begin{minipage}[t]{3.5cm}
    Center \\
    \end{minipage}
    &
    \begin{minipage}[t]{3.8cm}
    99.8\% \\
    \end{minipage}
    & 
    \begin{minipage}[t]{5.0cm}
    90.6\% ($\pm$1 pixel)\\
    100\% \ ($\pm$2 pixel)\\
    \end{minipage}
    \\ 
    \begin{minipage}[t]{3.5cm}
    Rotation \\
    \end{minipage}
    &
    \begin{minipage}[t]{3.8cm}
    97.8\% \\
    \end{minipage}
    & 
    \begin{minipage}[t]{5.0cm}
    98.6\% ($\pm$1$^\circ$)\\
    100\% \ ($\pm$2$^\circ$)\\
    \end{minipage}
    \\ 
    \begin{minipage}[t]{3.5cm}
    Thickness (13.6 mrad) \\
    \end{minipage}
    &
    \begin{minipage}[t]{3.8cm}
    99.1\% \\
    \end{minipage}
    & 
    \begin{minipage}[t]{5.0cm}
    81.0\% ($\pm$1 nm vs. LSF)\textsuperscript{*}\\
    95.7\% ($\pm$2 nm vs. LSF)\textsuperscript{*}\\
    96.5\% ($\pm$3 nm vs. LSF)\textsuperscript{*}\\
    \end{minipage}
    \\ 
    \begin{minipage}[t]{3.5cm}
    Tilt (13.6mrad) \\
    \end{minipage}
    &
    \begin{minipage}[t]{3.8cm}
    97.2\% \\
    \end{minipage}
    & 
    \begin{minipage}[t]{5cm}
    74.0\% ($\pm$0.25 mrad vs. LSF)\textsuperscript{*}\\
    93.7\% ($\pm$0.50 mrad vs. LSF)\textsuperscript{*}\\
    97.3\% ($\pm$0.71 mrad vs. LSF)\textsuperscript{*}\\
    \end{minipage}
    \\ 
    \begin{minipage}[t]{3.5cm}
    Thickness (19.1 mrad) \\
    \end{minipage}
    &
    \begin{minipage}[t]{3.8cm}
    98.0\% \\
    \end{minipage}
    & 
    \begin{minipage}[t]{5cm}
    91.0\% ($\pm$1 nm vs. LSF)\textsuperscript{*}\\
    93.6\% ($\pm$2 nm vs. LSF)\textsuperscript{*}\\
    94.2\% ($\pm$3 nm vs. LSF)\textsuperscript{*}\\
    \end{minipage}
    \\ 
    \begin{minipage}[t]{3.5cm}
    Tilt (19.1 mrad) \\
    \end{minipage}
    &
    \begin{minipage}[t]{3.8cm}
    91.3\% \\
    \end{minipage}
    & 
    \begin{minipage}[t]{5cm}
    61.5\% ($\pm$0.25 mrad vs. LSF)\textsuperscript{*}\\
    94.9\% ($\pm$0.50 mrad vs. LSF)\textsuperscript{*}\\
    100\% \ ($\pm$0.71 mrad vs. LSF)\textsuperscript{*}\\
    \end{minipage}
    \\
    \hline
    \multicolumn{3}{l}{\textsuperscript{*}\footnotesize{While LSF may not necessarily report the true value, all LSF results are confirmed by  visual inspection.}}
  \end{tabular}
  \label{table:testing}
\end{table} 


\begin{table}[h!]
	\caption{Typical PACBED thickness and tilt measurements from the convolutional neural networks with corresponding classification confidence.}
	\centering
  \centering
  \begin{tabular}{ | c | m{3.6cm} | m{3.6cm} | m{3cm} | }
    \hline
    \textbf{Experiment} & \textbf{CNN} & \textbf{LSF} & \textbf{Visual Matching} \\ \hline
    \begin{minipage}{.2\textwidth}
      \includegraphics[width=\linewidth, height=3.2cm]{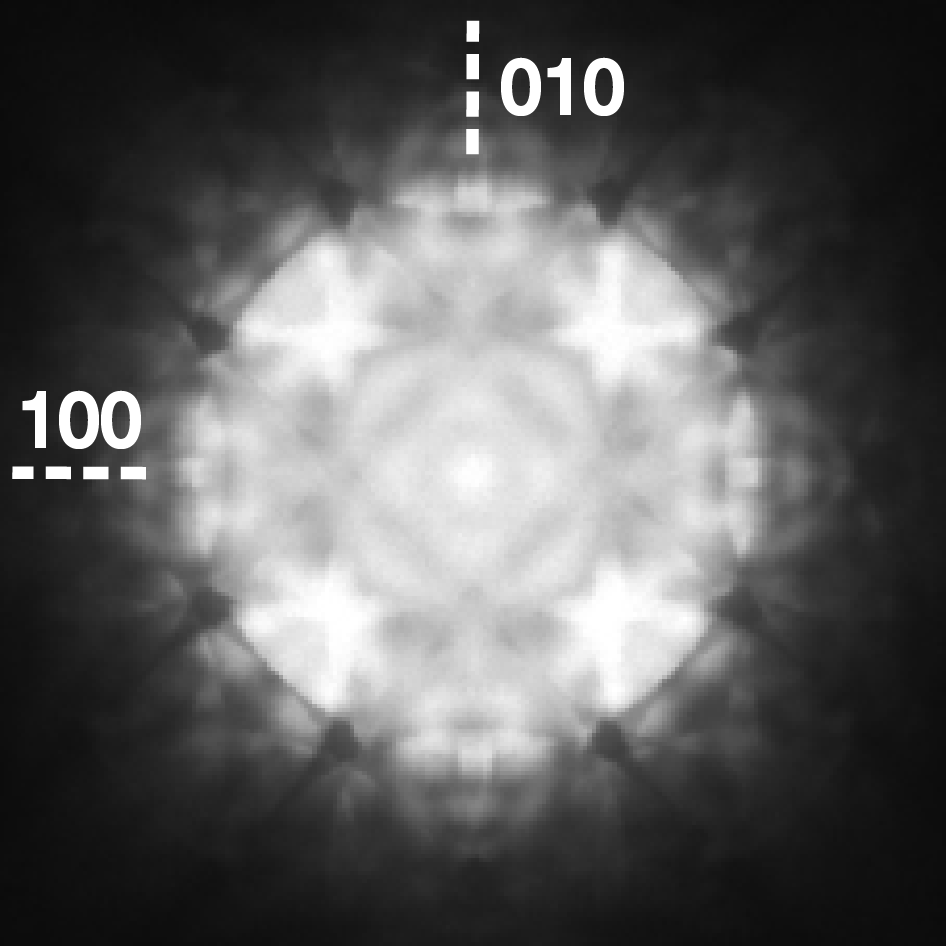}
    \end{minipage}
    &
    \begin{minipage}[t]{3.6cm}
    \textbf{Thickness:} \\
    85 nm 79.6\%  \\
    104 nm 17.8\%  \\
    \textbf{Tilt ([100]/[010]):} \\
    0/-0.25 mrad 96.8\% \\
    0/-0.5 mrad 2.8\% \\
    \end{minipage}
    & 
    \begin{minipage}[t]{3.6cm}
    \textbf{Thickness:} \\
    86 nm (+6/-6 nm)  \\ 
     \\
    \textbf{Tilt ([100]/[010]):} \\
    0.25/-0.25 mrad \\
    \\
    \end{minipage}
    &
    \begin{minipage}[t]{3cm}
    \textbf{Thickness:} \\
    85-86 nm  \\ \\ \\ \\ \\
    \end{minipage}
    \\ \hline
    \begin{minipage}{.2\textwidth}
      \includegraphics[width=\linewidth, height=3.2cm]{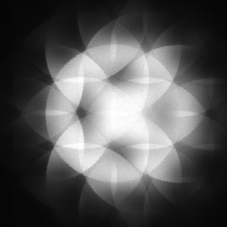}
    \end{minipage}
    &
    \begin{minipage}[t]{3.6cm}
    \textbf{Thickness:} \\
    22 nm 58.8\%  \\
    21 nm 41.1\%  \\
    \textbf{Tilt ([100]/[010]):} \\
    -2.5/0 mrad 80.4\% \\
    -2/-0.5 mrad 9.5\% \\
    \end{minipage}
    & 
    \begin{minipage}[t]{3.6cm}
    \textbf{Thickness:} \\
    22 nm (+2/-2 nm)  \\
    \\
    \textbf{Tilt ([100]/[010]):} \\
    -2.5/0 mrad\\
    \\ 
    \end{minipage}
    &
    \begin{minipage}[t]{3cm}
    \textbf{Thickness:} \\
    20-25 nm  \\ \\ \\ \\ \\
    \end{minipage}
    \\ \hline
    \begin{minipage}{.2\textwidth}
      \includegraphics[width=\linewidth, height=3.2cm]{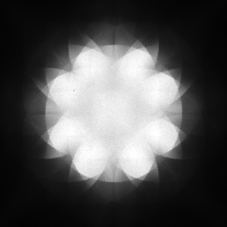}
    \end{minipage}
    &
    \begin{minipage}[t]{3.6cm}
    \textbf{Thickness:} \\
    28 nm 100\%  \\
    \\
    \textbf{Tilt ([100]/[010]):} \\
    0.75/1 mrad 99.9\% \\
    \\
    \end{minipage}
    & 
    \begin{minipage}[t]{3.6cm}
    \textbf{Thickness:} \\
    28 nm (+4/-4 nm)  \\
    \\
    \textbf{Tilt ([100]/[010]):} \\
    1/1.5 mrad \\
    \\ 
    \end{minipage}
    &
    \begin{minipage}[t]{3cm}
    \textbf{Thickness:} \\
    26-30 nm  \\ \\ \\ \\ \\
    \end{minipage}
    \\ \hline
    \begin{minipage}{.2\textwidth}
      \includegraphics[width=\linewidth, height=3.2cm]{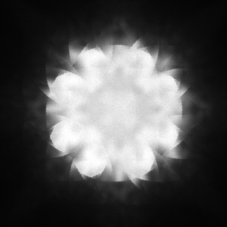}
    \end{minipage}
    &
    \begin{minipage}[t]{3.6cm}
    \textbf{Thickness:} \\
    38 nm 59.2\%  \\
    39 nm 40.8\%  \\
    \textbf{Tilt ([100]/[010]):} \\
    -1.5/1.5 mrad 52.8\% \\
    -1/1.5 mrad 44.6\% \\
    \end{minipage}
    & 
    \begin{minipage}[t]{3.6cm}
    \textbf{Thickness:} \\
    39 nm (+4/-3 nm)  \\
    \\
    \textbf{Tilt ([100]/[010]):} \\
    -1.5/1.5 mrad \\
    \\ 
    \end{minipage}
    &
    \begin{minipage}[t]{3cm}
    \textbf{Thickness:} \\
    37-39 nm  \\ \\ \\ \\ \\
    \end{minipage}
    \\
    \hline
  \end{tabular}
  \label{table:exp_compare}
\end{table}

The first neural networks -- size, shift, and rotation -- achieve near 100\% accuracy within $\pm$2\%, $\pm$2 pixels, and $\pm$2$^{\circ}$, respectively. Supplementary Video 1 highlights this procedure for a set of raw PACBED patterns directly from experiment, where the CNNs are able to center the patterns and match their size. Beyond simple cases, the patterns are well-aligned even when the zero order disk is not well defined, i.e.~for thick sample regions or when there is significant sample tilt. The accuracy of these automated measurements guarantees robust pattern alignment for the subsequent thickness and tilt classification. 


In comparing the time required for processing, the trained CNNs vastly outperform the LSF method. The time for LSF to determine thickness and tilt is significant, and in some cases is more than 30 min/pattern. In contrast, an average rate of 0.1 s/pattern is achieved for the CNNs, as measured from the raw image input to the output. For experiment  thickness and tilt measurements, the CNN approach exhibits a high degree of accuracy. Using LSF as a benchmark, near 95.7\% CNN measurements are matched within $\pm$2 nm and 96.5\% within $\pm$3 nm for 13.6 mrad patterns. Similarly, 19.1 mrad CNN results match LSF 93.6\% and 94.2\% within $\pm$2 nm and $\pm$3 nm, respectively. For tilt measurements, the CNN 93.7\%(97.3\%) matches within 0.50(0.71 mrad) tilt for 13.6 mrad, and 94.9\% (100\%) matches within 0.50 (0.71 mrad) tilt for 19.1 mrad.  It is important to note, however, that the results from LSF should be regarded as a benchmark reference, and not necessarily the absolute true values. In particular, the LSF approach can report incorrect thicknesses in contaminated areas or when the pattern quality is low.


\section{Factors Influencing Network Performance}

\subsection{Thickness Dependence}
\label{sec:thicknessDependence}

\begin{figure}[h!]
\centering
\includegraphics[width=5.54in]{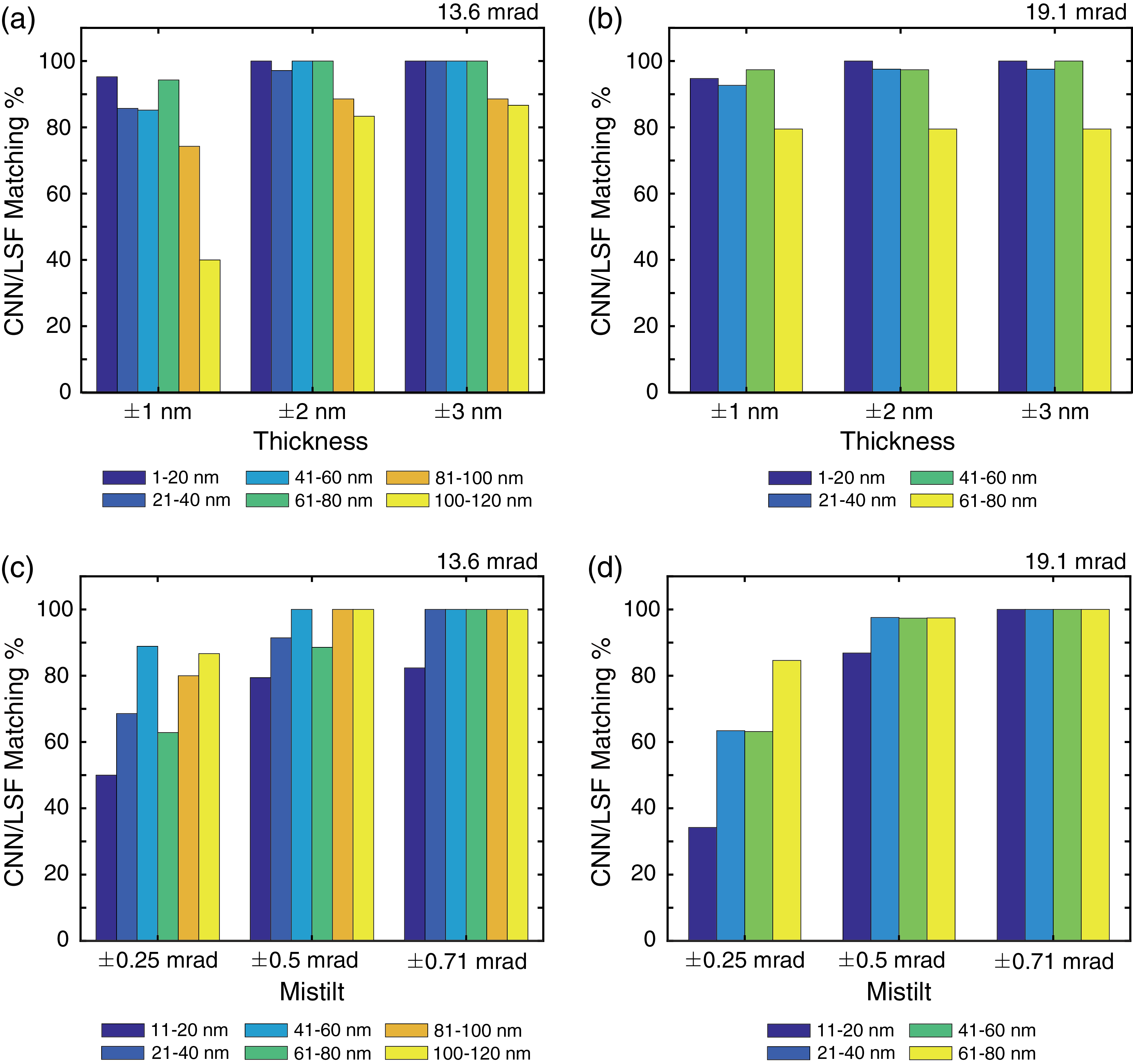}
\caption{Percentage of CNN and LSF thickness and tilt measurements matching as a function of specimen thickness: (a, c) 13.6 mrad and (b, d) 19.1 mrad}
\label{fig:hist_error}
\end{figure}

The CNN thickness measurement precision is found to be dependent upon sample thickness, as shown in Figure \ref{fig:hist_error}. The CNN exhibits the best performance for thicknesses below 60 nm, where over 95\% of the patterns match within $\pm$2 nm of the LSF method (Figure \ref{fig:hist_error}a,b). The agreement between the CNN and LSF decreases with increasing thickness, with the poorest performance occuring within the 100-120 nm range.

\begin{figure}[h!]
\centering
\includegraphics[width=3.34in]{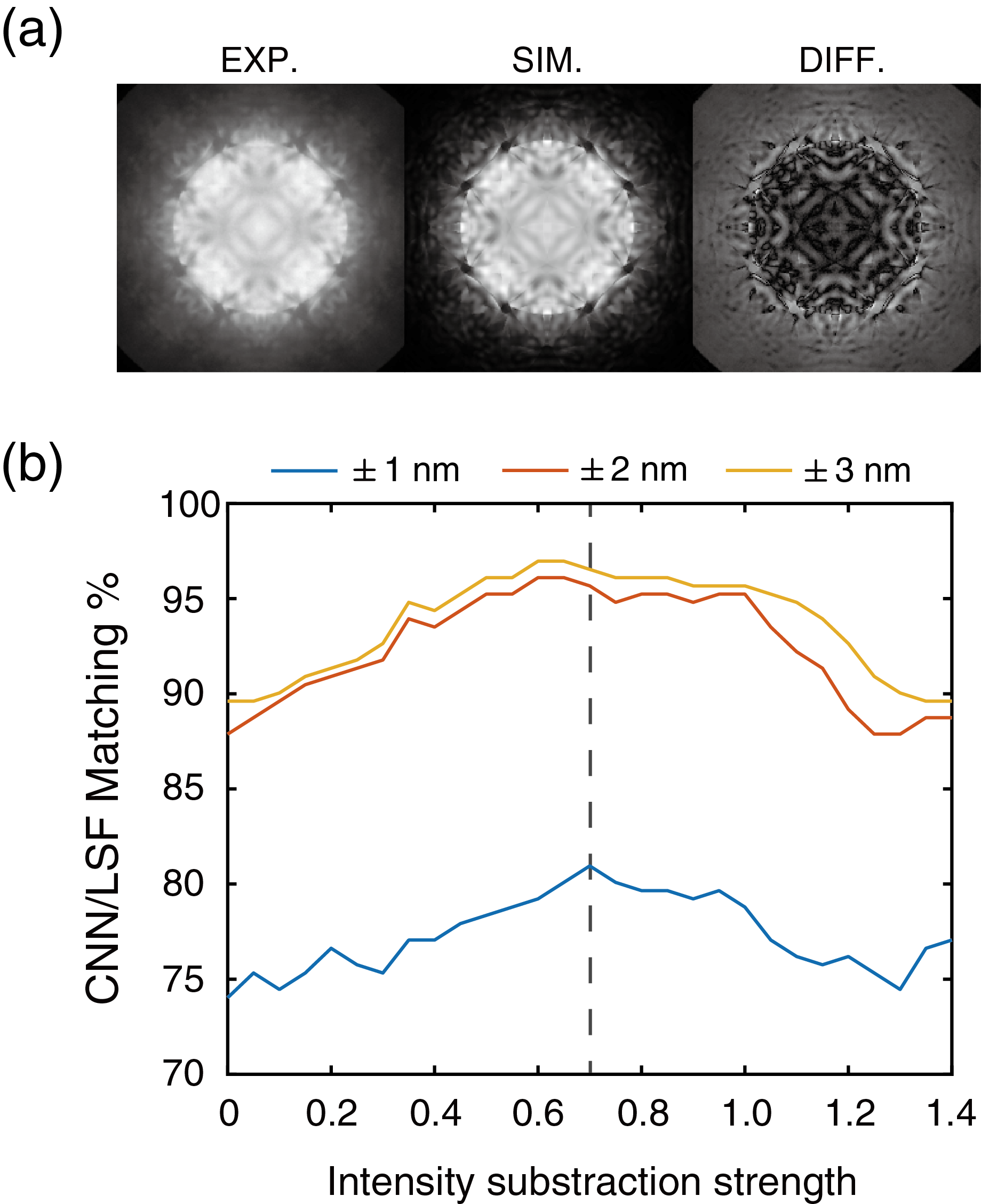}
\caption{Percentage of CNN and LSF thickness measurement that match  as function of uniform background intensity subtraction strength. Abscissa values are on the same intensity scale as the image.)}
\label{fig:int_comp}
\end{figure}

Close inspection of experiment and simulations suggests that the error for thicker regions is most likely due to the pattern discrepancy. As seen in Figure \ref{fig:int_comp}a for example, a 100 nm pattern from experiment shows additional background intensity at the periphery of the pattern and blurring. In addition, a discrepancy is also observed in the region of overlapped disks as shown in the difference map. These differences arise largely from inelastic thermal diffuse and plasmon scattering \cite{Allen_muSTEM,Egerton}.  To minimize the effect of inelastic scattering on the CNN classification, a uniform background subtraction is applied, with the value found through iteration within the network. This is shown in Figure \ref{fig:int_comp}b, where the accuracy of PACBED thickness determination can be optimized with appropriate background subtraction. For the case of SrTiO$_3$, 0.7/nm and 0.3/nm are optimal for 13.6 mrad and 19.1 mrad, respectively. 


\begin{figure}[h!]
\centering
\includegraphics[width=5.54in]{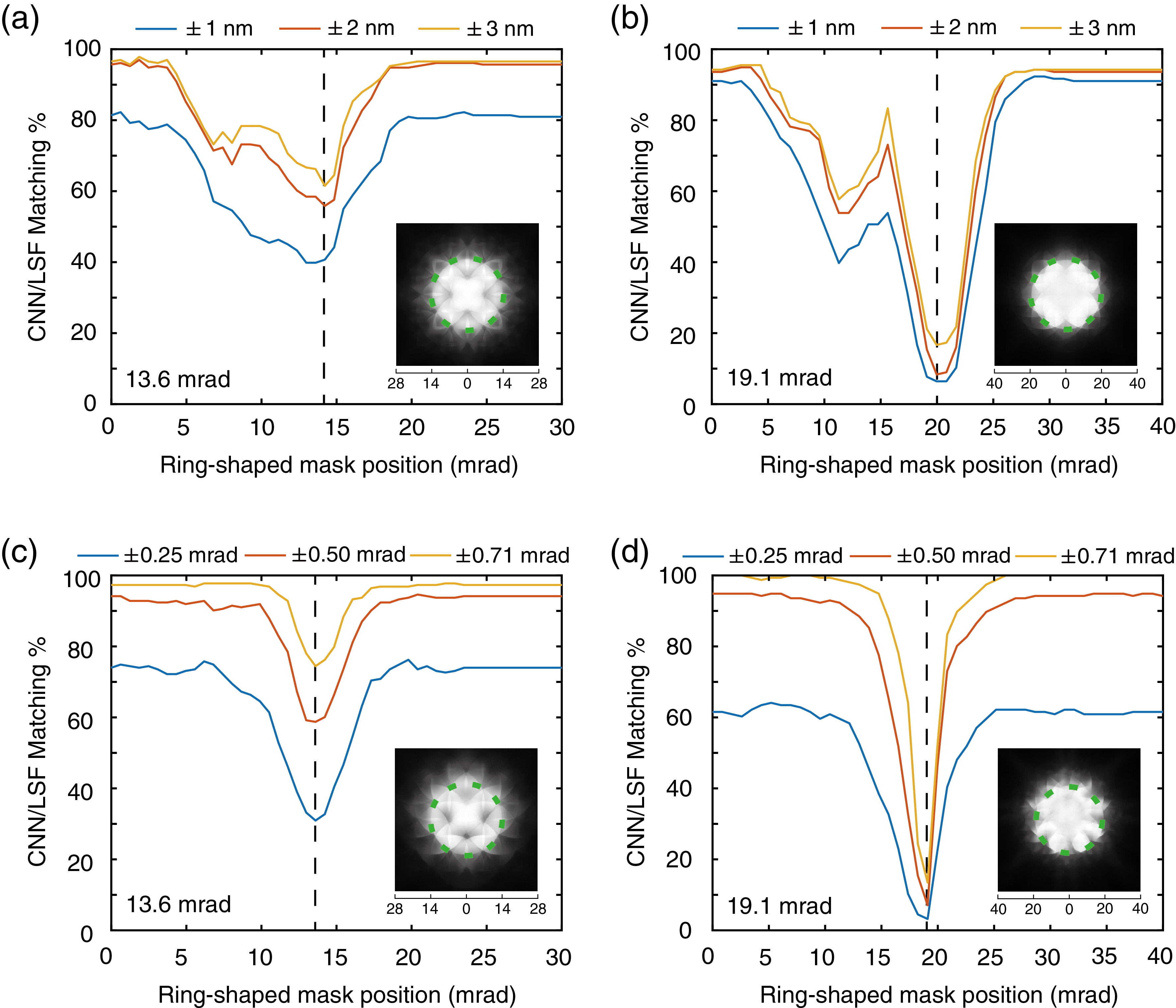}
\caption{Response of CNN ability to classify patterns when removing pattern features via the band-pass mask for thickness measurements with (a) 13.6 mrad and (b) 19.1 mrad and  tilt measurements with (c) 13.6 mrad and (d) 19.1 mrad}
\label{fig:band-pass_mask}
\end{figure}

Sensitivity of the CNNs to pattern features is explored further in Figures \ref{fig:band-pass_mask}a,b, where the CNN/LSF match rate as a function of band-pass mask position is provided. The performance degradation occurs at about 18 mrad and 24 mrad for the 13.6 and 19.1 convergence semi-angle, respectively. As the band-pass mask is placed closer to the PACBED center, the agreement between CNN and LSF decreases to a minimum roughly at the convergence semi-angle. A similar trend can be seen in all the percentage of matching CNN/LSF results within $\pm1$, $\pm2$, and $\pm3$ nm. This suggests that the periphery  around the central disk up to about this limit is heavily weighted by the CNN for thickness determination.  

\begin{figure}[h!]
\center
\includegraphics[width=3.54in]{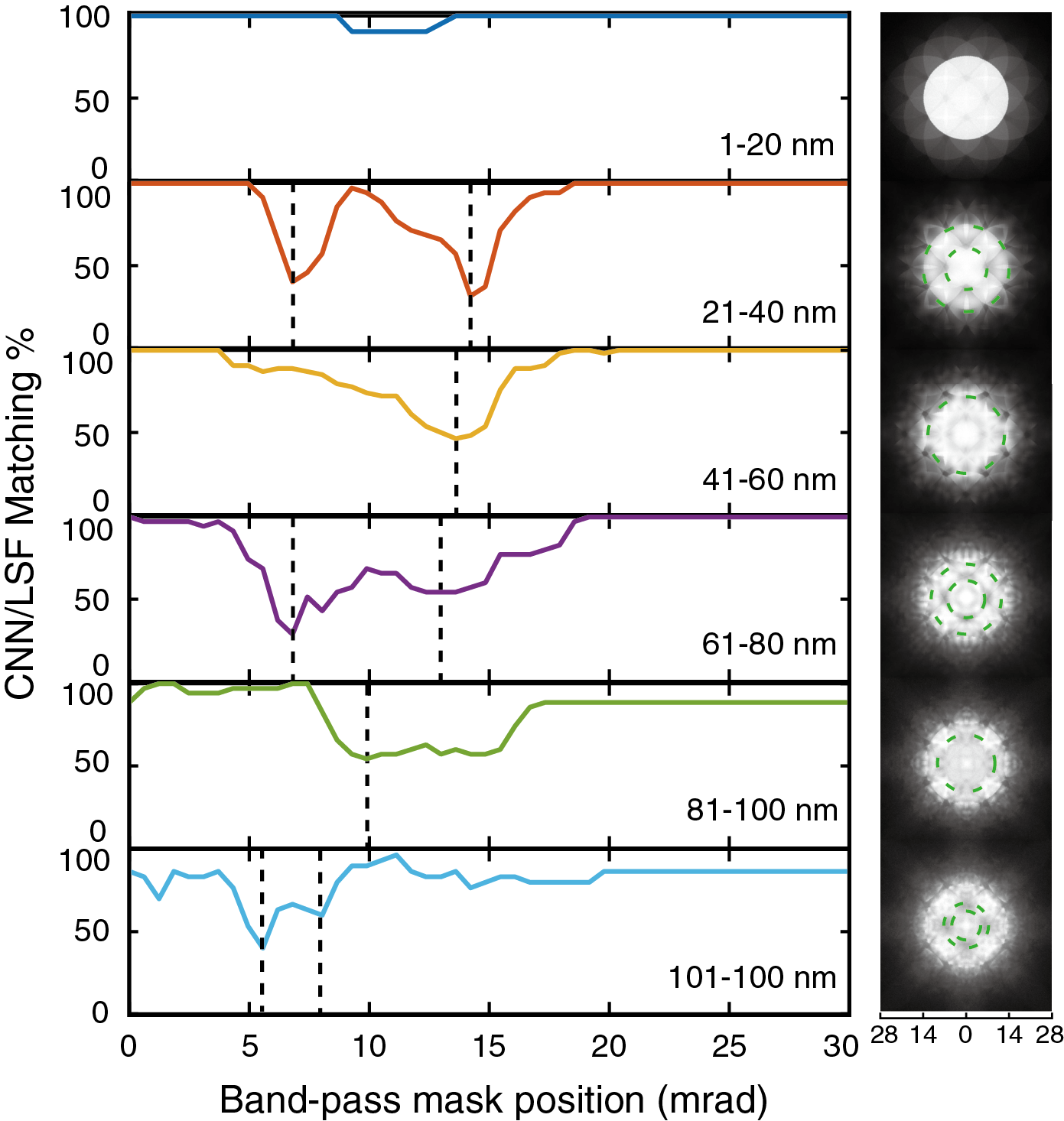}
\caption{Response of CNN ability to classify patterns for indicated specimen thickness ranges. The range of band-pass mask positions that lead to local minima are indicated by corresponding dashed lines (left) and circles (right).} 
\label{fig:band-pass_mask_range}
\end{figure}

The band-pass analysis also shows that the CNN precision is closely linked to the individual thickness ranges. As highlighted in Figure \ref{fig:band-pass_mask_range}, the percentage CNN/LSF results that match is minimized when the regions with the greatest brightness and contrast are excluded from the patterns. This demonstrates that the CNN automatically identified the most relevant local pattern features. For the thickness range of 1-20 nm, however, the CNN performance is nearly flat because the patterns themselves do not exhibit strong features within this regime.  More importantly, the reduction of the CNN accuracy occurs for a broad range of band-pass masks. This suggests the CNN utilizes multiple regions of the PACBED patterns why classifying, which leads to increased robustness.


\subsection{Tilt Dependence}
\label{sec:tiltdepend}
The CNN agreement with LSF tends to improve with increasing thickness due to the increasing pattern detail, as shown in Figures \ref{fig:hist_error}c-d. The CNN classification uncertainty within $\pm$0.25 mrad, however, is significant, which limits tilt precision to about $\pm$0.5 mrad. For specimen thicknesses below $\leq$10 nm, the CNN is only reliable to within about $\pm$1 mrad. It should be noted here that the LSF method also struggles to identify tilt within this thickness range, where incorrect measurements are also frequently observed. 

For 13.6 mrad tilt measurements, the greatest CNN response to the occurs with a band-pass mask with a starting semi-angle of 10-18 mrad. The minimum is located at about 13.5 mrad, which is near the edges of zero order disk as marked by the dashed circle in the PACBED pattern in Figure \ref{fig:band-pass_mask}b. A ~20-30\% drop is found for the percentage of CNN/LSF results matching within $\pm$0.25 mrad, $\pm$0.50 mrad, and $\pm$ 0.71 mrad. It is also important to note that for tilt measurements the influence of the band-pass mask is relatively small compared with thickness. At the central disk periphery, the CNN performance does not change when pattern features are removed with the mask. This indicates that the CNN for tilt largely relies on pattern features outside of the central pattern, and thus it is essential to include these features when capturing the pattern with the detector.


\subsection{Convergence semi-angle dependence}

 

\begin{figure*}[h!]
\centering
\includegraphics[width=4.45in]{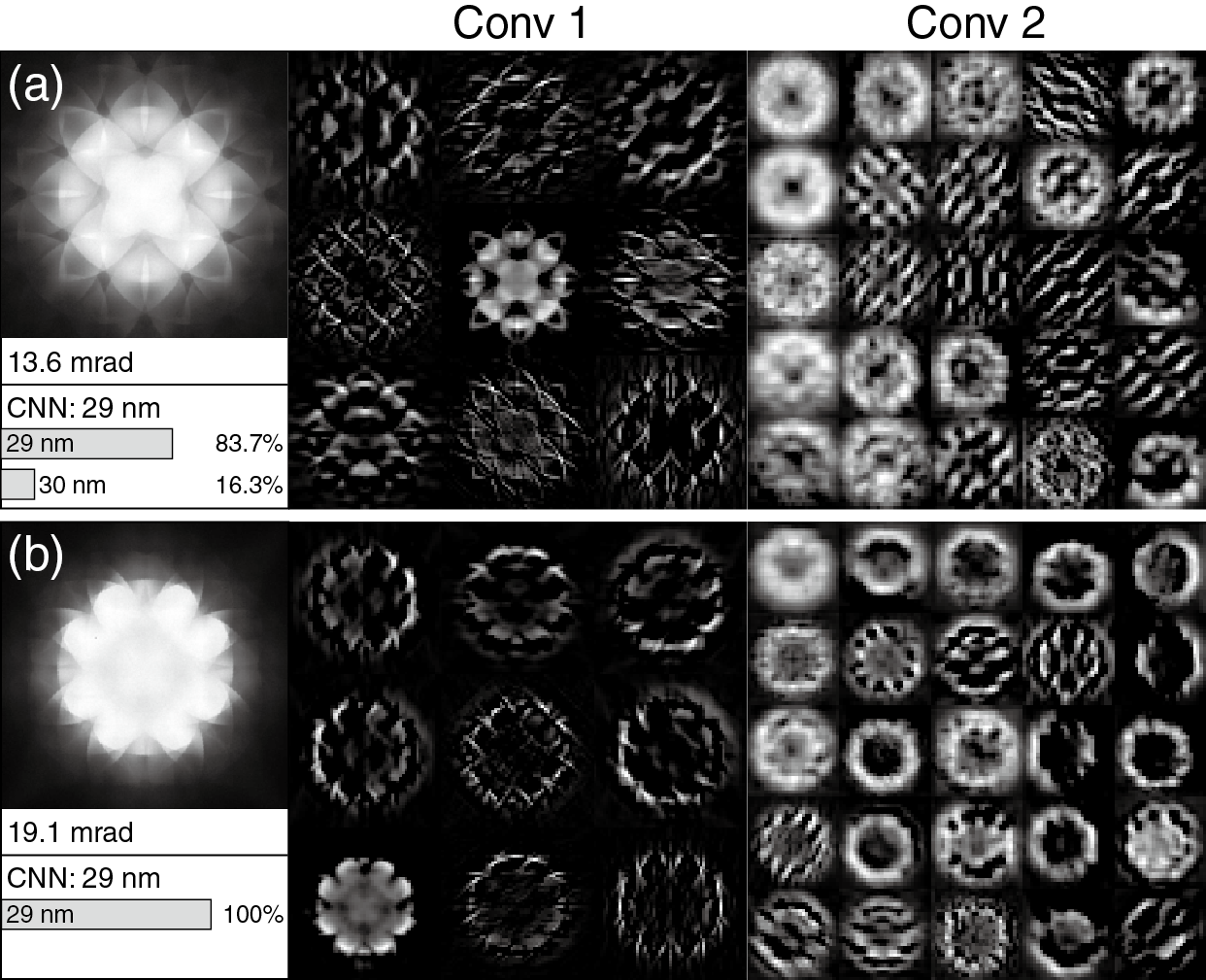}
\caption{Thickness measurement neural network activation in the first and second convolutional layers.}
\label{fig:activation}
\end{figure*}

The probe forming convergence semi-angle is a critical parameter for the neural network classification as it controls the degree of diffraction disk overlap and pattern detail. To better understand the effect of convergence semi-angle on the CNN performance, the network activation is visualized in Figure \ref{fig:activation} for both the 13.6 and 19.1 mrad networks. A variety of bright and dark patterns are observed for the 13.6 mrad CNN, where the strongest activation occurs both inside the zero order disk and at its edge. The activation for 19.1 mrad, however, is more localized at the zero-order disk edge. The localized region of network activation and response of 19.1 mrad is attributed to the reduction of pattern features when the convergence semi-angle is large. As a result, the 19.1 mrad CNN trained for determining thickness cannot handle thicker sample measurement, which has also been reported for LSF methods \cite{Pollock201786}. In order to reduce the complexity of the network, the acceptable thickness range is reduced to 1-80 nm for SrTiO$_3$, which improves the network reliability (87-92.3\%) when compared to LSF. 



\subsection{Dose Dependence}

\begin{figure}[h!]
\centering
\includegraphics[width=3.24in]{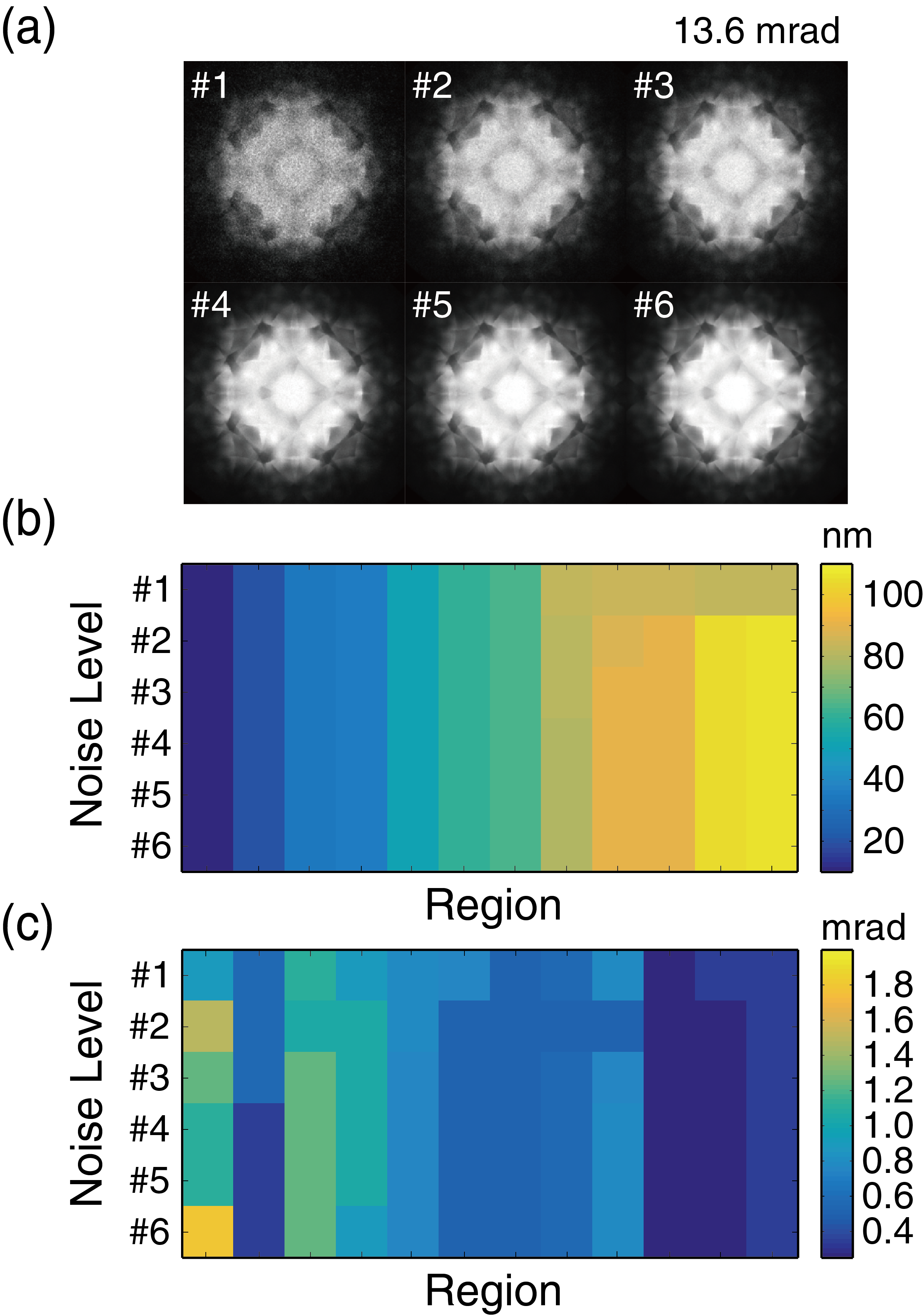}
\caption{Effect of noise level on 13.6 mrad CNN thickness and tilt measurements. (a) Example patterns from one specimen region at different dose levels: \#1: 4.5$\times$10$^6$, \#2: 2.3$\times$10$^7$, \#3: 4.7$\times$10$^7$, \#4: 2.3$\times$10$^8$, \#5: 4.7$\times$10$^8$, \#6: 9.3$\times$10$^8$ electrons/e$^-$. (b) and (c) CNN measurement of thickness and tilt at various thicknesses and dose. }
\label{fig:noise}
\end{figure}

The PACBED pattern clarity is affected by electron dose. Figure \ref{fig:noise}a, for example, shows patterns from the same area, but with a dose ranging from 4.5$\times$10$^6$ to 9.3$\times$10$^8$ e$^-$/pattern. While the pattern alignment is robust for all dose levels, the impact on the thickness measurements is shown in Figure \ref{fig:noise}b. When the dose is $\geq$ 2.3$\times$10$^7$ e$^-$/pattern, the CNN reports the same thicknesses. Similar results are observed for both 13.6 and 19 mrad patterns. Below this dose, the thickness error increases. A similar dose effect is also seen in the CNN tilt measurement, but the dose threshold for robust measurements is around $\sim$4.7$\times$10$^7$ e$^-$/pattern. This dose sensitivity for tilt measurements is due to tilt determination being heavily weighted by feature changes on the pattern periphery as highlighted in Section \ref{sec:tiltdepend}.

\section{Observations \& Opportunities}

\subsection{Application to 4D STEM datasets}

The CNN processing speed is well suited for PACBED quantification of big, 4D STEM datasets. As an example, Figure \ref{fig:2D_CNN} shows the result of thickness and tilt measurements over a 60$\times$60 nm$^2$ area. See Supplementary Video 2 for the collection of PACBED patterns from this region exhibiting changes in both thickness and sample distortion. It is important to emphasize that PACBED analysis from this region is challenging due to the combination of a thickness gradient with local crystal distortion.  

\begin{figure*}[h!]
\centering
\includegraphics[width=5.54in]{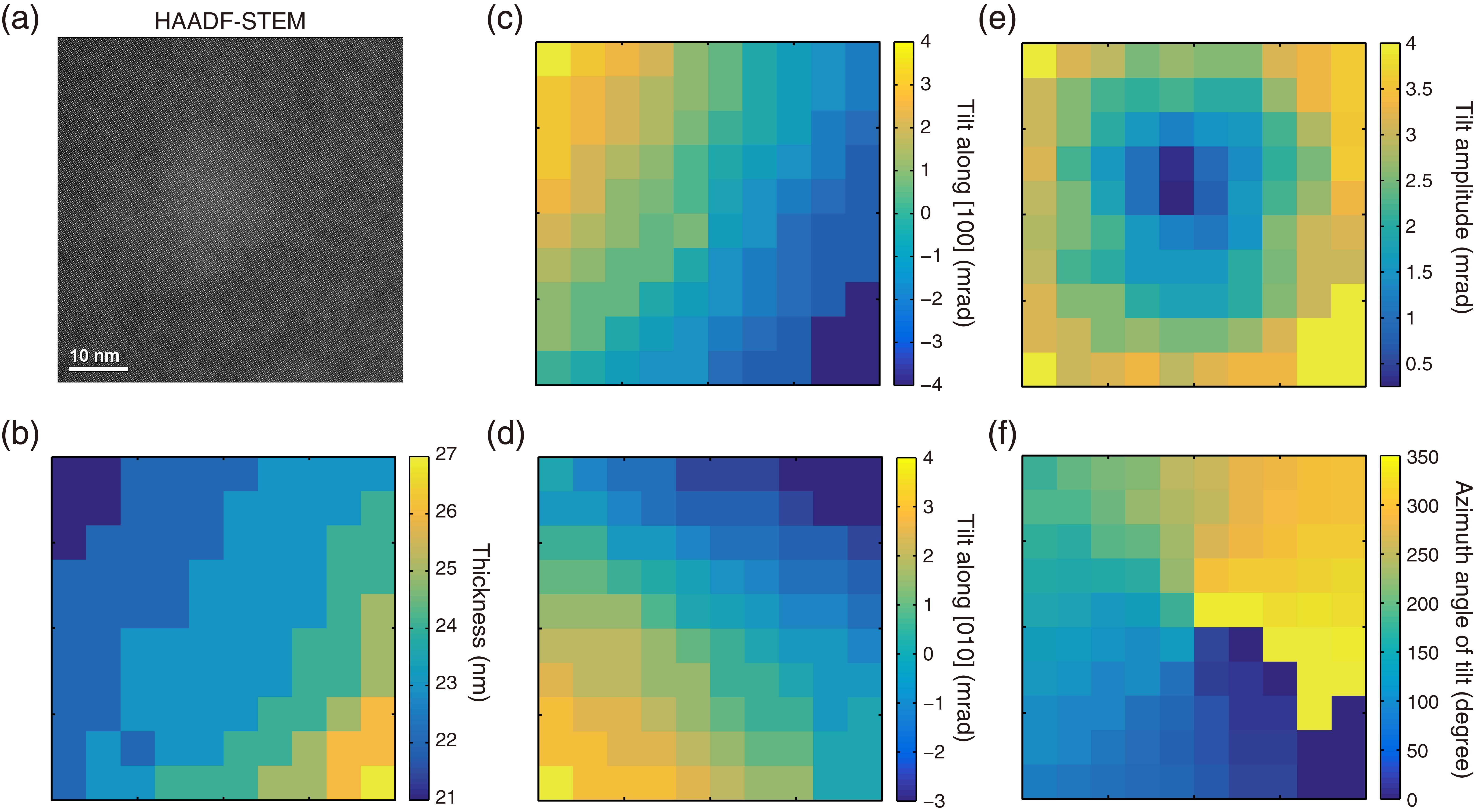}
\caption{(a) HAADF STEM image of a local 60$\times$60 nm$^2$ area and corresponding local (b) thickness and tilt along (c) $\left[001\right]$ and (d) $\left[100\right]$. The tilt magnitude and the azimuthal angle  are shown in (e) and (f), respectively.}
\label{fig:2D_CNN}
\end{figure*}

The CNN identified thicknesses are presented in Figure \ref{fig:2D_CNN}b, which shows the smooth, wedge shape of the specimen from the top left to the bottom right.  As shown, the thickness changes by about 5 nm across the analysis region with an average thickness of 23 nm. This is particularly important when quantifying STEM images where a 5 nm difference in thickness introduces a 25\% change in the image signal \cite{LeBeau_2008_PRL, LeBeau_2008_UM, LeBeau_nanoLett}.

In addition to thickness, the crystal is found to be distorted by $\pm$4 mrad along [100] and [010] directions as in Figures \ref{fig:2D_CNN}c-d. Using this information, the tilt amplitude and azimuth are reported in Figure \ref{fig:2D_CNN}e-f, with a maximum value of 4.2 mrad. Tilts of this magnitude are more than sufficient to alter electron channeling conditions, and significantly modify signal contrast \cite{Maccagnano-Zacher2008} or shift positions of atom columns in STEM \cite{Brown201776, Zhou2016110}. Furthermore, the tilt map reveals that the local region is bowl shaped. This curvature can, for example, also critically influence absorption for quantitative EDS. With the applied automated process, these factors can then be readily taken into account in further quantitative analysis.

\subsection{Generality of the approach}

While the neural network is trained for automatically aligning 13.6 mrad and 19.1 mrad datasets, the neural network is sufficiently generalized to align patterns acquired at other convergence semi-angles. A test using simulated patterns shows that the current CNN can align PACBED pattern size and shift for convergence semi-angles greater than 13 mrad when captured at 200 kV, see Supplementary Video 3. Furthermore, the neural networks achieve good generalization for other directions and different structures even though they are trained to align $\left[001\right]$ SrTiO$_3$ patterns. For example, the CNNs align $\left[011\right]$ PbMg$_{1/3}$Nb$_{2/3}$O$_3$ (also a perovskite) patterns without additional training. In fact, the pattern rotation CNN aligns $\left[100\right]$ and $\left[0\bar{1}1\right]$ along the pattern horizontal and vertical, respectively.  This is particularly fortuitous, as the $\left[011\right]$ exhibits 2-fold symmetry, whereas the training set is 4-fold. Thickness and tilt of the aligned pattern can then be  estimated using the SrTiO$_3$ datasets, but precise and accurate measurements require additional training.

\subsection{Hybrid CNN+LSF Analysis}
\label{hybrid}

\begin{figure*}[h!]
\centering
\includegraphics[width=6.0in]{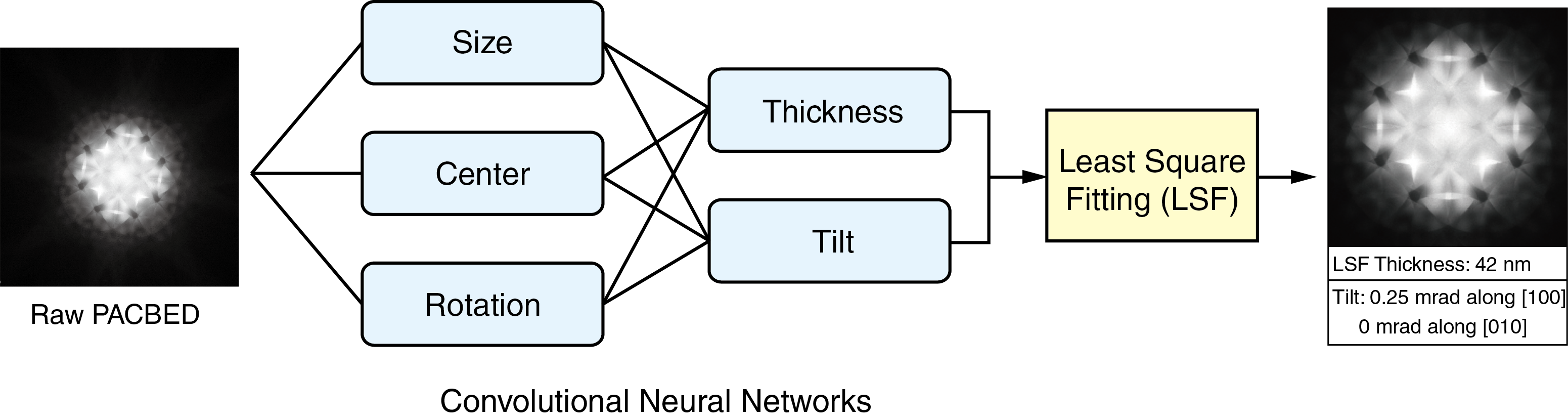}
\caption{Hybrid CNN+LSF architecture for PACBED measurement.}
\label{fig:hybrid}
\end{figure*}

The limitations outlined in the introduction of the LSF approach  can be overcome through a hybrid approach with CNN methods. As illustrated in Figure \ref{fig:hybrid}, the CNN approach developed here can support LSF via its automatic PACBED pattern alignment and thickness/tilt measurements as initial parameters, e.g.~the search range is then  limited  within $\pm$3 nm and $\pm$1 mrad for thickness and tilt respectively. This significantly narrows the LSF search range for a dramatically decreased time to find the global minimum. For that range, the average processing speed can be reduced by a factor of $>$ 100.  This is particularly useful for PACBED analysis where LSF may be essential and/or is the path of least resistance.  For example, this includes polarity determination \cite{Xiahan_APL}, electrostatic field quantification \cite{TAPLIN201669}, or quantifying octahedral distortion \cite{Jinwoo_2012, Jinwoo_2013_PRB, Zhang_2013}

\section{Conclusions}

A convolutional neural network approach has been developed to automatically measure key parameters from PACBED patterns. This includes the zero-order disk size, pattern center, rotation, thickness, and tilt. The network has been successfully transfer-trained using thousands of simulated patterns that are augmented with additional variables to account for random geometric distortion, size, shift, noise, and intensity variations. The trained networks show excellent accuracy for measuring PACBED patterns when compared with brute force methods. Through GPU acceleration, an overall processing rate of 0.1 s per pattern has been achieved, enabling fast analysis of 4D STEM data. Furthermore, a methodology has been developed to explore how the CNN responds to features across the PACBED patterns.  Overall, the approach provides a critical demonstration of how neural networks can be successfully implemented for the automated analysis of electron diffraction data.

\section*{Acknowledgments}

The authors gratefully acknowledge the Air Force Office of Scientific Research (FA9550-14-1-0182) for support of this research. We thank Dr. Rohan Dhall for discussion. This work was performed in part at the Analytical Instrumentation Facility (AIF) at North Carolina State University, which is supported by the State of North Carolina and the National Science Foundation (award number ECCS-1542015). The AIF is a member of the North Carolina Research Triangle Nanotechnology Network (RTNN), a site in the National Nanotechnology Coordinated Infrastructure (NNCI).

\section*{References}
\bibliographystyle{elsarticle-num}

\bibliography{refs.bib}

\end{document}